\documentclass[twocolumn,aps,superscriptaddress,showpacs]{revtex4}
\usepackage{amssymb}
\usepackage{amsmath}
\usepackage{graphicx}
\usepackage[normalem]{ulem}
\usepackage[dvips]{color}

\renewcommand\sout{\bgroup \color{red} \ULdepth=-.5ex \ULset}

\begin{document}

\title{Locating the inner edge of neutron star crust using terrestrial
nuclear laboratory data}
\author{Jun Xu}
\affiliation{Institute of Theoretical Physics, Shanghai Jiao Tong University, Shanghai
200240, China}
\author{Lie-Wen Chen}
\affiliation{Institute of Theoretical Physics, Shanghai Jiao Tong
University, Shanghai 200240, China} \affiliation{Center of
Theoretical Nuclear Physics, National Laboratory of Heavy-Ion
Accelerator, Lanzhou 730000, China}
\author{Bao-An Li}
\affiliation{Department of Physics, Texas A\&M University-Commerce, Commerce, TX
75429-3011, USA}
\author{Hong-Ru Ma}
\affiliation{Institute of Theoretical Physics, Shanghai Jiao Tong University, Shanghai
200240, China}

\begin{abstract}
Within both dynamical and thermodynamical approaches using the
equation of state for neutron-rich nuclear matter constrained by the
recent isospin diffusion data from heavy-ion reactions in the same
sub-saturation density range as the neutron star crust, the density
and pressure at the inner edge separating the liquid core from the
solid crust of neutron stars are determined to be $0.040$ fm$^{-3}$
$\leq \rho _{t}\leq 0.065$ fm$^{-3}$ and $0.01$ MeV/fm$^{3}$ $\leq
P_{t}\leq 0.26$ MeV/fm$^{3}$, respectively. These together with the
observed minimum crustal fraction of the total moment of inertia
allow us to set a new limit for the radius of the Vela pulsar
significantly different from the previous estimate. It is further
shown that the widely used parabolic approximation to the equation
of state of asymmetric nuclear matter leads systematically to
significantly higher core-crust transition densities and pressures,
especially with stiffer symmetry energy functionals.

\end{abstract}

\maketitle

\section{Introduction}
Having been the major testing grounds of our knowledge on the nature
of matter under extreme conditions, neutron stars are among the most
mysterious objects in the universe. To understand their structures
and properties has long been a very challenging task for both the
astrophysics and the nuclear physics community~\cite{Lat04}.
Theoretically, neutron stars are expected to have a solid inner
crust surrounding a liquid core. Knowledge on properties of the
crust plays an important role in understanding many astrophysical
observations~\cite%
{BPS71,BBP71,Pet95a,Pet95b,Lat00,Lat07,Ste05,Lin99,Hor04,Bur06,Owe05}. The
inner crust spans the region from the neutron drip-out point to the inner
edge separating the solid crust from the homogeneous liquid core. While the
neutron drip-out density $\rho _{out}$ is relatively well determined to be
about $4\times 10^{11}$ g/cm$^{3}$ \cite{Rus06}, the transition density $%
\rho _{t}$ at the inner edge is still largely uncertain mainly
because of our very limited knowledge on the equation of state
(EOS), especially the density dependence of the symmetry energy, of
neutron-rich nucleonic matter~\cite{Lat00,Lat07}. These
uncertainties have hampered our accurate understanding of many
important properties of neutron stars~\cite{Lat04,Lat00,Lat07}.

Recently, significant progress has been made in constraining the EOS
of neutron-rich nuclear matter using terrestrial laboratory
experiments (See Ref.~\cite{LCK08} for the most recent review). In
particular, the analysis of isospin-diffusion data
\cite{Tsa04,Che05a,LiBA05c} in heavy-ion collisions has constrained
tightly the density dependence of the symmetry energy in exactly the
same sub-saturation density region around the expected inner edge of
neutron star crust. Moreover, the obtained constraint on the
symmetry energy was found to agree with isoscaling analyses in
heavy-ion collisions \cite{She07}, the isotopic dependence of the
giant monopole resonance in even-A Sn isotopes \cite{LiT07}, and the
neutron-skin thickness of $^{208}$Pb \cite{Ste05b,LiBA05c,Che05b}.
In this paper, using the equation of state for neutron-rich nuclear
matter constrained by the recent isospin diffusion data from
heavy-ion reactions in the same sub-saturation density range as the
neutron star crust, we determine the inner edge of neutron star
crusts. Consequently, the limit on the radius of the Vela pulsar is
significantly different from the previous estimate. In addition, we
find that the widely used parabolic approximation (PA) to the EOS of
asymmetric nuclear matter enhances significantly the transition
densities and pressures, especially with stiffer symmetry energy
functionals.

\section{The theoretical method}
The inner edge corresponds to the phase transition from the
homogeneous matter at high densities to the inhomogeneous matter at
low densities. In principle, the inner edge can be located by
comparing in detail relevant properties of the nonuniform solid
crust and the uniform liquid core mainly consisting of neutrons,
protons and electrons ($npe$ matter). However, this is practically
very difficult since the inner crust may contain nuclei having very
complicated geometries, usually known as the `nuclear
pasta'~\cite{Lat04,Rav83,Oya93,Hor04,Ste08}. Furthermore, the
core-crust transition is thought to be a very weak first-order phase
transition and model calculations lead to a very small density
discontinuities at the transition~\cite{Pet95b,Dou00,Dou01,Hor03}.
In practice, therefore, a good approximation is to search for the
density at which the uniform liquid first becomes unstable against
small amplitude density fluctuations with clusterization. This
approximation has been shown to produce very small error for the
actual core-crust transition density and it would yield the exact
transition density for a second-order phase
transition~\cite{Pet95b,Dou00,Dou01,Hor03}. Several such methods
including the dynamical
method~\cite{BPS71,BBP71,Pet95a,Pet95b,Dou00,Oya07,Duc07}, the
thermodynamical method~\cite{Kub07,Lat07,Wor08} and the random phase
approximation (RPA)~\cite{Hor01,Hor03} have been applied extensively
in the literature. Here, we use both the dynamical and
thermodynamical methods.

In the dynamical method, the stability condition of a homogeneous
$npe$ matter against small periodic density perturbations with
clusterization can be well approximated by \cite%
{BPS71,BBP71,Pet95a,Pet95b,Oya07}
\begin{equation}
V_{dyn}(k)=V_{0}+\beta k^{2}+\frac{4\pi e^{2}}{k^{2}+k_{TF}^{2}}>0,
\label{Vdyn}
\end{equation}%
where
\begin{eqnarray}
V_{0} &=&\frac{\partial \mu _{p}}{\partial \rho _{p}}-\frac{(\partial \mu
_{n}/\partial \rho _{p})^{2}}{\partial \mu _{n}/\partial \rho _{n}},\text{ }%
k_{TF}^{2}=\frac{4\pi e^{2}}{\partial \mu _{e}/\partial \rho _{e}},  \notag \\
\beta  &=&D_{pp}+2D_{np}\zeta +D_{nn}\zeta ^{2},~~\zeta
=-\frac{\partial \mu _{n}/\partial \rho _{p}}{\partial \mu
_{n}/\partial \rho _{n}},  \notag
\end{eqnarray}%
and $k$ is the wavevector of the spatially periodic density
perturbations and $\mu _{i}$ is the chemical potential of particle
$i$. In the above expressions, we used the relation $\frac{\partial
\mu _{n}}{\partial \rho _{p}}=\frac{\partial \mu _{p}}{\partial \rho
_{n}}$ following $\frac{\partial \mu _{n}}{\partial \rho _{p}}=\frac{\partial }{%
\partial \rho _{p}}\left( \frac{\partial \varepsilon }{\partial \rho _{n}}%
\right) =\frac{\partial }{\partial \rho _{n}}\left( \frac{\partial
\varepsilon }{\partial \rho _{p}}\right) =\frac{\partial \mu
_{p}}{\partial \rho _{n}}$ with $\varepsilon $ being the energy
density of the $npe$ matter. The three terms in Eq.~(\ref{Vdyn})
represent, respectively, the contributions from the bulk nuclear
matter, the density gradient (surface) terms and the Coulomb
interaction. For the coefficients of density gradient terms
we use their empirical values of $D_{pp}=D_{nn}=D_{np}=132$ MeV$\cdot $fm$%
^{5}$ consistent with the Skyrme-Hartree-Fock calculations~\cite%
{Oya07,XCLM08}. At $k_{\min }=[(\frac{4\pi e^{2}}{\beta }%
)^{1/2}-k_{TF}^{2}]^{1/2}$, the $V_{dyn}(k)$ has the minimal value of $%
V_{dyn}(k_{\min })=V_{0}+2(4\pi e^{2}\beta )^{1/2}-\beta k_{TF}^{2}$~\cite%
{BPS71,BBP71,Pet95a,Pet95b,Oya07}. Its vanishing point determines
the $\rho _{t}$.

The thermodynamical method requires the system to obey the intrinsic
stability condition~\cite{Cal85} or the following
inequalities~\cite{Kub07,Lat07}
\begin{equation}\label{ther1}
-\left(\frac{\partial P}{\partial v}\right)_\mu>0,
\end{equation}
\begin{equation}\label{ther2}
-\left(\frac{\partial \mu}{\partial q_c}\right)_v>0.
\end{equation}
These conditions are equivalent to require the convexity of the
energy per particle in the single phase~\cite{Kub07,Lat07} by
ignoring the finite size effects due to surface and Coulomb energies
as shown in the following. In the above, the $P=P_b+P_e$ is the
total pressure of the $npe$ system with the contributions $P_b$ and
$P_e$ from baryons and electrons, respectively. The $v$ and $q_c$
are the volume and charge per baryon number. The $\mu$ is the
chemical potential defined as
\begin{equation}
\mu=\mu_n-\mu_p.
\end{equation}
In fact, Eq.~(\ref{ther1}) is simply the well-known mechanical
stability condition of the system at a fixed $\mu$. It ensures that
any local density fluctuation will not diverge. On the other hand,
Eq.~(\ref{ther2}) is the charge or chemical stability condition of
the system at a fixed density. It means that any local charge
variation violating the charge neutrality condition will not
diverge. If the $\beta$-equilibrium condition is satisfied, namely
$\mu=\mu_e$, the electron contribution to the pressure $P_e$ is only
a function of the chemical potential $\mu$, and in this case one can
rewrite Eq.~(\ref{ther1}) as
\begin{equation}
-\left(\frac{\partial P_b}{\partial v}\right)_\mu>0.
\end{equation}
By using the relation $\frac{\partial E(\rho,x_p)}{\partial x_p} =
-\mu$, one can get~\cite{Kub07}
\begin{eqnarray}
-\left( \frac{\partial P}{\partial v}\right) _{\mu } &=&2\rho ^{3}\frac{%
\partial E(\rho ,x_{p})}{\partial \rho }+\rho ^{4}\frac{\partial ^{2}E(\rho
,x_{p})}{\partial \rho ^{2}}  \notag \\
&-&\rho ^{4}\left( \frac{\partial ^{2}E(\rho ,x_{p})}{\partial \rho
\partial x_{p}}\right) ^{2}/\frac{\partial ^{2}E(\rho
,x_{p})}{\partial x_{p}^{2}}>0 \label{Vther}
\end{eqnarray}
\begin{equation}\label{ther4}
-\left(\frac{\partial q_c}{\partial \mu}\right)_v =
1/\frac{\partial^2 E(\rho,x_p)}{\partial x_p^2} + \frac{\partial
\rho_e}{\partial \mu_e}/\rho,
\end{equation}
where $q_c=x_p-\rho_e/\rho$. The $\rho=1/v$ is the baryon density
and the $E(\rho,x_p)$ is the energy per baryon for the nucleonic
matter. Within the free Fermi gas model, the density of electrons
$\rho_e$ is uniquely determined by the electron chemical potential
$\mu_e$. Then the thermodynamical relations Eq.~(\ref{ther1}) and
Eq.~(\ref{ther2}) are identical to~\citep{Lat07,Kub07}
\begin{widetext}
\begin{eqnarray}\label{ther5}
-\left(\frac{\partial P_b}{\partial v}\right)_\mu = \rho^2 \left[2
\rho \frac{\partial E(\rho,x_p)}{\partial \rho} + \rho^2
\frac{\partial^2 E(\rho,x_p)}{\partial \rho^2} -
\left(\frac{\partial^2 E(\rho,x_p)}{\partial \rho
\partial x_p}\rho\right)^2/\frac{\partial^2 E(\rho,x_p)}{\partial x_p^2}\right]>0,
\end{eqnarray}
\begin{equation}\label{ther6}
-\left(\frac{\partial q_c}{\partial \mu}\right)_v =
1/\frac{\partial^2 E(\rho,x_p)}{\partial
x_p^2}+\frac{\mu^2_e}{\pi^2\hbar^3\rho}>0,
\end{equation}
\end{widetext}
respectively. The second inequality is usually valid. Thus, the
following condition from the first one
\begin{eqnarray} \label{Vther}
V_{ther} &=& 2 \rho \frac{\partial E(\rho,x_p)}{\partial \rho} +
\rho^2 \frac{\partial^2 E(\rho,x_p)}{\partial \rho^2} \notag \\
&-&\left(\frac{\partial^2 E(\rho,x_p)}{\partial \rho
\partial x_p}\rho\right)^2/\frac{\partial^2 E(\rho,x_p)}{\partial
x_p^2}>0
\end{eqnarray}
determines the thermodynamical instability region.

Based on general thermodynamic relations and $\frac{\partial \mu
_{n}}{\partial \rho _{p}}=\frac{\partial \mu _{p}}{\partial \rho
_{n}}$, one can show~\cite{XCLM08}
\begin{equation}
\frac{2}{\rho }\frac{\partial E}{\partial \rho }\frac{\partial ^{2}E}{%
\partial x_{p}^{2}}+\frac{\partial ^{2}E}{\partial \rho ^{2}}\frac{\partial
^{2}E}{\partial x_{p}^{2}}-\left( \frac{\partial ^{2}E}{\partial \rho
\partial x_{p}}\right) ^{2}=\frac{\partial \mu _{n}}{\partial \rho _{n}}%
\frac{\partial \mu _{p}}{\partial \rho _{p}}-\left( \frac{\partial \mu _{n}}{%
\partial \rho _{p}}\right) ^{2}.  \label{relation}
\end{equation}%
Thus, for $\partial ^{2}E/\partial x_{p}^{2}>0$, Eq. (\ref{Vther})
is equivalent to requiring a positive bulk term $V_{0}$ in
Eq.~(\ref{Vdyn}). Generally speaking, the $\rho_t$ is in the
subsaturation density region where $\partial ^{2}E/\partial
x_{p}^{2}$ is almost always positive for all models. Therefore, the
thermodynamical stability condition is simply the limit of the
dynamical one as $k\rightarrow 0$ by neglecting the Coulomb
interaction.

To locate the inner edge of neutron star crust, we use the same MDI
(Momentum Dependent Interaction) EOS that was used in analyzing the
isospin diffusion in heavy-ion reactions~\cite{Che05a,LiBA05c}. The
MDI interaction is based on a modified finite-range Gogny effective
interaction \cite{Das03} and has been extensively used in our
previous work \cite{LCK08}. The baryon potential
energy density part of the MDI EOS can be expressed as~\cite%
{Das03,Che05a,Che07}
\begin{eqnarray}
V(\rho ,T,\delta ) &=&\frac{A_{u}\rho _{n}\rho _{p}}{\rho _{0}}+\frac{A_{l}}{
2\rho _{0}}(\rho _{n}^{2}+\rho _{p}^{2})+\frac{B}{\sigma +1}\frac{\rho
^{\sigma +1}}{\rho _{0}^{\sigma }}  \notag \\
&\times &(1-x\delta ^{2})+\frac{1}{\rho _{0}}\sum_{\tau ,\tau ^{\prime
}}C_{\tau ,\tau ^{\prime }}  \notag \\
&\times &\int \int d^{3}pd^{3}p^{\prime }\frac{f_{\tau }(\vec{r},\vec{p}
)f_{\tau ^{\prime }}(\vec{r},\vec{p}^{\prime })}{1+(\vec{p}-\vec{p}^{\prime
})^{2}/\Lambda ^{2}}.  \label{MDIVB}
\end{eqnarray}
Here $\tau$ is $1/2$ ($-1/2$) for neutrons (protons) and $\delta
=1-2x_{p}$ is the isospin asymmetry. The meaning and values of other
parameters can be found in Refs.~\cite{Das03,Che05a}. The parameter
$x$ is introduced to vary only the density dependence of the
symmetry energy while keeping other properties of the nuclear EOS
fixed \cite{Che05a}. In particular, the
symmetry energy $E_{sym}(\rho )=\frac{1}{2}\left( \frac{\partial ^{2}E}{%
\partial \delta ^{2}} \right) _{\delta =0}$ at saturation density $\rho
_{0}=0.16$ fm$^{-3}$ is fixed at $30.54$ MeV for all values of the
parameter $x$. The isospin symmetric part of the MDI EOS was shown
to agree with the experimental constraints obtained from
relativistic heavy-ion collisions up to about $5$ times the
saturation density~\cite{Kra08}.

\section{Results}
\begin{figure}[t!]
\centering
\includegraphics[scale=0.8]{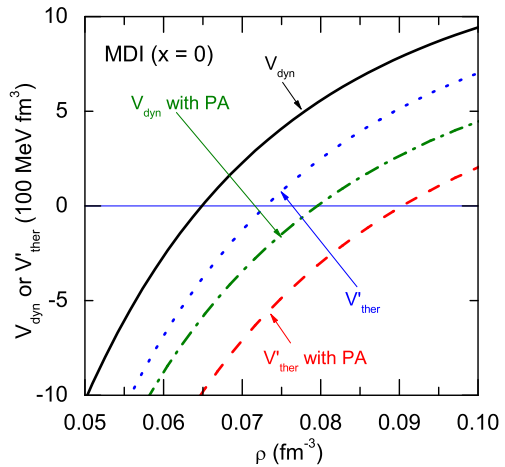}
\caption{{\protect\small (Color online) The density dependence of
$V_{dyn}$ and $V^\prime_{ther}$ for MDI interaction with $x=0$ using
both the dynamical and thermodynamical methods with the full EOS and
its parabolic approximation (PA).}} \label{VrhoMDI}
\end{figure}

Firstly, to show how the uniform npe matter becomes stable from
unstable with increasing baryon density and how to locate the
transition density and see the difference between the dynamical and
thermodynamical methods as well as effects of the PA, we show in
Fig.~\ref{VrhoMDI} the density dependence of $V_{dyn}$ and
$V^\prime_{ther}$ using the MDI interaction with $x=0$ within both
the dynamical and thermodynamical methods with the full EOS and its
PA. Here, we have defined
\begin{equation}
V^\prime_{ther}=V_{ther} \frac{\partial^2 E}{\partial x_p^2}
/\left(\rho^2 \frac{\partial \mu_n}{\partial \rho_n}\right)
\end{equation}
and it should be noted that $V^\prime_{ther}$ has the same vanishing
point as the $V_{ther}$ and the same dimension as the $V_{dyn}$. For
the MDI interaction with $x=0$ the transition densities using the
full EOS within the dynamical and thermodynamical method are $0.065$
fm$^{-3}$ and $0.073$ fm$^{-3}$, respectively. While the
corresponding results using the PA are $0.080$ fm$^{-3}$ and $0.090$
fm$^{-3}$, respectively. Thus, the transition densities are
generally lower with the dynamical method as the density gradient
term and the Coulomb interaction make the system more stable.
However, the PA significantly lifts the transition density
regardless of the approach used. In fact, the difference between
calculations using the full EOS and its PA is much larger than that
caused by using the two different methods.

\begin{figure}[tbh]
\includegraphics[scale=0.8]{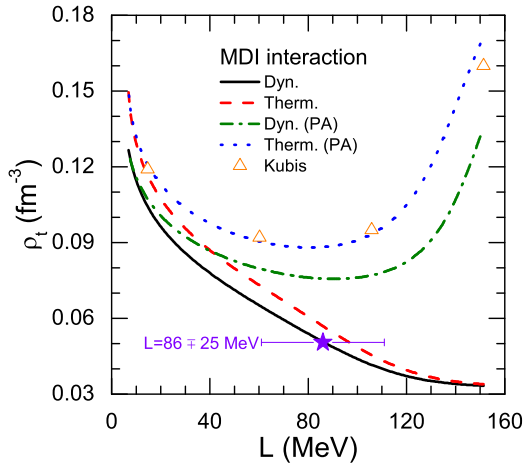}
\caption{{\protect\small (Color online) The }$\protect%
\rho _{t}$ {\protect\small as a function of $L$ from the dynamical
and thermodynamical methods with and without the parabolic approximation
in the MDI interaction. The triangles are obtained by
Kubis~\protect\cite{Kub07} and the star with error bar represents
}${\protect\small L=86}\pm {\protect\small 25}${\protect\small \
MeV.}} \label{rhotL}
\end{figure}
Shown in Fig.~\ref{rhotL} is the $\rho _{t}$ as a
function of the slope parameter of the symmetry energy $L=3\rho _{0}\frac{%
\partial E_{sym}(\rho )}{\partial \rho }|_{\rho =\rho _{0}}$ with the MDI
interaction. For comparisons, we have included results using both
the dynamical and thermodynamical methods with the full EOS and its
parabolic approximation, i.e., $E(\rho ,\delta )=E(\rho ,\delta
=0)+E_{sym}(\rho )\delta ^{2}+O(\delta^{4})$ from the same MDI
interaction. With the full MDI EOS, it is clearly seen that the
$\rho _{t}$ decreases almost linearly with increasing $L$ for both
methods. This feature is consistent with the RPA
results~\cite{Hor01}. It is interesting to see that both the
dynamical and thermodynamical methods give very similar results with
the former giving slightly smaller $\rho _{t}$ than the later (the
difference is actually less than $0.01$ fm$^{-3}$) and this is due
to the fact that the former includes the density gradient and
Coulomb terms which make the system more stable and lower the
transition density. The small difference between the two methods
implies that the effects of density gradient terms and Coulomb term
are unimportant in determining the $\rho _{t}$. On the other hand,
surprisingly, the PA drastically changes the results, especially for
stiffer symmetry energies (larger $L$ values). Also included in
Fig.~\ref{rhotL} are the predictions by Kubis using the PA of the
MDI EOS in the thermodynamical approach~\cite{Kub07}. Furthermore,
we find that in the parabolic approximation, using the EOS of
$E(\rho ,\delta )=E(\rho ,\delta =0)+[E(\rho ,\delta =1)-E(\rho
,\delta =0)]\delta ^{2}+O(\delta^{4})$ leads to almost the same
results. The large error introduced by the PA is understandable
since the $\beta $-stable $npe$ matter is usually highly
neutron-rich and the contribution from the higher order terms in
$\delta $ is appreciable. This is especially the case for the
stiffer symmetry energy which generally leads to a more neutron-rich
$npe $ matter at subsaturation densities. In addition, simply
because of the energy curvatures involved in the stability
conditions, the contributions from higher order terms in the EOS are
multiplied by a larger factor than the quadratic term. These
features agree with the early finding \cite{Arp72} that the $\rho
_{t}$ is very sensitive to the fine details of the nuclear EOS. We
notice that the EOS of asymmetric nuclear matter always contains the
higher-order terms in isospin asymmetry (at least for the kinetic
part of the EOS). Our results indicate that one may introduce a huge
error by assuming {\it a priori} that the EOS is parabolic for a
given interaction in calculating the $\rho _{t}$. We thus apply the
experimentally constrained $L$ to the $\rho_t-L$ correlation
obtained using the full EOS in constraining the $\rho _{t}$. In the
following, we will mainly focus on the dynamical method as it is
more complete and realistic.

The transport model analysis of the isospin diffusion data from
heavy-ion collisions allowed us to constrain the parameter $x$ in
Eq.~(\ref{MDIVB}) to be between $x=0$ and $x=-1$ in the density
range of $0.3\rho_0$ and $1.2\rho_0$~\cite{Che05a,LiBA05c}. With the
full MDI EOS, the slope parameter was determined to be $L=86\pm 25$
MeV. The approximately $30\%$ error in $L$ is systematic in nature
mainly due to the uncertainty of the in-medium nucleon-nucleon cross
sections used in the transport model~\cite{LiBA05c}. The statistical
errors in both the data analysis~\cite{Tsa04} and model
calculations~\cite{Che05a,LiBA05c} are less than the systematic
error. The error in $L$ will lead to roughly similar systematic
errors in all quantities we study here. As shown in
Fig.~\ref{rhotL}, the constrained $L$ then limits the transition
density to $0.040$ fm$^{-3} $ $\leq \rho _{t}\leq 0.065$ fm$^{-3}$.
It is interesting to mention that both approaches used here for
finding the core-crust transition density in the $npe$ matter in
neutron stars have also been widely used in studying the spinodal
decomposition density associated with the liquid-gas phase
transition in nuclear matter, see, e.g.,
Refs.~\cite{Mul95,LiKo97,Mar03,Xu08} for some recent applications in
neutron-rich matter. As expected, see, e.g., Ref.~\cite{Sie83}, the
two phase transitions are asymptotically but inherently related.
Applying both the dynamical and thermodynamical approaches to
symmetric nuclear matter (SNM) at zero temperature for MDI
interaction by ignoring the Coulomb and surface terms, we find a
transition density of $0.63\rho_0$. It is consistent with the
spinodal decomposition density for SNM at zero temperature from the
relativistic mean field model~\cite{Mul95} and Skyrme density
functionals~\cite{Mar03}. On the other hand, recent
studies~\cite{Mar07,Rio08} indicate that some effects beyond the
mean-field approximation may affect the EOS of asymmetric nuclear
matter in the low-density region. Thus, it would be interesting to
study how the effects beyond the mean-field approximation as well as
other effects such as the many-body forces may change the core-crust
transition density of neutron stars.

\begin{figure}[tbh]
\includegraphics[scale=0.84]{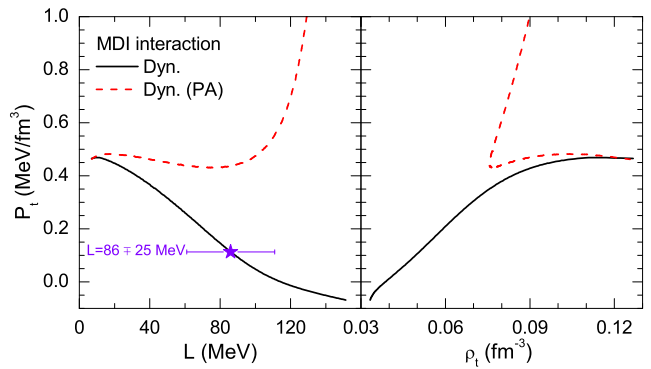}
\caption{{\protect\small (Color online) The $P_{t}$ as functions of $L$ and $%
\protect\rho _{t}$ by using the dynamical method with and without parabolic
approximation in the MDI interaction. The star with error bar in the left
panel represents }${\protect\small L=86}\pm {\protect\small 25}$
{\protect\small \ MeV.}}
\label{PtLrhot}
\end{figure}

The pressure at the inner edge, $P_{t}$, is also an important quantity which
might be measurable indirectly from observations of pulsar glitches \cite%
{Lat07,Lin99}. Shown in Fig.~\ref{PtLrhot} is the $P_{t}$ as
functions of $L$ and $\rho _{t}$ by using the dynamical method with
both the full MDI EOS and its PA. Again, it is seen that the PA
leads to huge errors for larger (smaller) $L$ ($\rho _{t}$) values.
For the full MDI EOS, the $P_{t}$ decreases (increases) with the
increasing $L$ ($\rho _{t}$) while it displays a complex relation
with $L$ or $\rho _{t}$ for the PA. The complex behaviors are due to
the fact that the $\rho _{t}$ does not vary monotonically with $L$
for the PA as shown in Fig.~\ref{rhotL}. From the constrained $L$
values, the $P_{t}$ is limited between $0.01$ MeV/fm$^{3}$ and
$0.26$ MeV/fm$^{3}$.

\begin{figure}[tbh]
\includegraphics[scale=0.8]{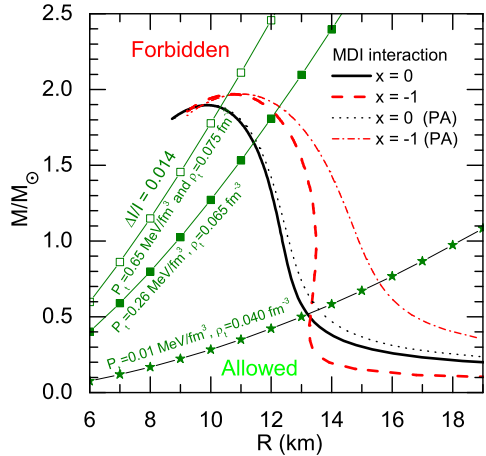}
\caption{(Color online) The $M$-$R$ relation of static neutron stars
from the full EOS and its parabolic approximation in the MDI
interaction with $x=0$ and $x=-1$. For the Vela pulsar, the
constraint of $\Delta I/I>0.014$ implies that allowed masses and
radii lie to the right of the line linked with solid squares
($\protect\rho _{t}=0.065$ fm$^{-3}$ and $P_{t}=0.26$ MeV/fm$^{3}$,
the upper bounds obtained in the present work), solid stars
($\protect\rho _{t}=0.040$ fm$^{-3}$ and $P_{t}=0.01$ MeV/fm$^{3}$,
the lower bounds obtained in the present work) or open squares
($\protect\rho _{t}=0.075$ fm$^{-3}$ and $P_{t}=0.65$ MeV/fm$^{3}$,
used in Ref.~\protect\cite{Lin99}).} \label{MR}
\end{figure}

The constrained values of $\rho _{t}$ and $P_{t}$ have important
implications on many properties of neutron
stars~\cite{Pet95a,Lat07,Hor04,Oya07}. As an example, here we
examine their impact on constraining the mass-radius ($M$-$R$)
correlation of neutron stars. The crustal fraction of the total
moment of inertia $\Delta I/I$ can be well approximated
by~\cite{Lin99,Lat00,Lat07}
\begin{eqnarray}
\frac{\Delta I}{I} &\approx&\frac{28\pi
P_{t}R^{3}}{3Mc^{2}}\frac{(1-1.67\xi
-0.6\xi ^{2})}{\xi }  \notag \\
&\times &\left[ 1+\frac{2P_{t}(1+5\xi -14\xi ^{2})}{\rho _{t}m_{b}c^{2}\xi
^{2}}\right] ^{-1},  \label{dI}
\end{eqnarray}%
where $m_{b}$ is the mass of baryons and $\xi =GM/Rc^{2}$ with $G$
being the gravitational constant. As it was stressed in
Ref.~\cite{Lat00}, the $\Delta I/I$ depends sensitively on the
symmetry energy at subsaturation densities through the $P_t$ and
$\rho_t$, but there is no explicit dependence upon the
higher-density EOS. So far, the only known limit of $\Delta
I/I>0.014$ was extracted from studying the glitches of the Vela
pulsar \cite{Lin99}. This together with the upper bounds on the
$P_t$ and $\rho_t$ ($\rho _{t}=0.065$ fm$^{-3}$ and $%
P_{t}=0.26$ MeV/fm$^{3}$) sets approximately a minimum radius of
$R\geq 4.7+4.0M/M_{\odot }$ km for the Vela pulsar. The radius of
the Vela pulsar is predicted to exceed $10.5$ km should it have a
mass of $1.4M_{\odot }$. A more restrictive constraint will be
obtained from the lower bounds of $\rho _{t}=0.040$ fm$^{-3}$
($P_{t}=0.01$ MeV/fm$^{3}$) which is indicated by the curve with
solid stars in Fig.~\ref{MR} and it can be approximately
parameterized by $R=5.5+14.5M/M_{\odot }$ km. It is thus seen that
the error of the transition density and pressure obtained in the
present work is still large and it leads to large uncertainties for
the mass-radius relation of the Vela pulsar. As a conservative
estimate, we thus deduce a constraint of $R\geq 4.7+4.0M/M_{\odot }$
km using the upper bounds on the $P_t$ and $\rho_t$ obtained in the
present work. We notice that a constraint of $R\geq
3.6+3.9M/M_{\odot }$ km for this pulsar has previously been derived
in Ref.~\cite{Lin99} by using $\rho _{t}=0.075$ fm$^{-3}$ and
$P_{t}=0.65$ MeV/fm$^{3}$. However, the constraint obtained in the
present work using for the first time data from both the terrestrial
laboratory experiments and astrophysical observations is
significantly different and actually it is more stringent.

To put the above constraints on the Vela pulsar in perspective, we
show them in Fig.~\ref{MR} together with the M$-$R relation by
solving the Tolman-Oppenheimer-Volkoff (TOV) equation. In the
latter, we use the well-known BPS EOS \cite{BPS71} for the outer
crust. In the inner crust with $\rho _{out}<\rho <\rho _{t},$ the
EOS is largely uncertain and following Ref. \cite{Hor03}, we use an
EOS of the form $P=a+b\epsilon ^{4/3}$ with the constants $a$ and
$b$ determined by the total pressure $P$ and total energy density
$\epsilon$ at $\rho_{out}$ and $\rho _{t}$. The full MDI EOS and its
parabolic approximation with $x=0$ and $x=-1$ are used for the
uniform liquid core with $\rho \geq \rho _{t}$. In this way, the $P$
is a continuous function of the $\epsilon $ at the boundary between
different regions as required. Assuming that the core consists of
only the $npe$ matter without possible new degrees of freedom or
phase transitions at high densities, the PA leads to a larger radius
for a fixed mass compared to the full MDI EOS. Furthermore, using
the full MDI EOS with $x=0$ and $x=-1$ constrained by the heavy-ion
reaction experiments, the radius of a canonical neutron star of
$1.4M_{\odot }$ is tightly constrained within $11.9$ km to $13.2$
km, which is consistent with the constraint $R\geq 4.7+4.0M/M_{\odot
}$ km for the Vela pulsar.

\section{Summary}
In summary, the density and pressure at the inner edge separating
the liquid core from the solid crust of neutron stars are determined
to be $0.040$ fm$^{-3}$ $\leq \rho _{t}\leq 0.065$ fm$^{-3}$ and
$0.01$ MeV/fm$^{3}$ $\leq P_{t}\leq 0.26$ MeV/fm$^{3}$,
respectively, using the MDI EOS of neutron-rich nuclear matter
constrained by the recent isospin diffusion data from heavy-ion
reactions in the same sub-saturation density range as the neutron
star crust. These constraints allow us to set a new limit on the
radius of the Vela pulsar. It is significantly different from the
previous estimate and thus puts a new constraint for the mass-radius
relation of neutron stars. Furthermore, we find that the widely used
parabolic approximation to the EOS of asymmetric nuclear matter
leads systematically to significantly higher core-crust transition
densities and pressures, especially for the energy density
functional with stiffer symmetry energies. Our results thus indicate
that one may introduce a huge error by assuming {\it a priori} that
the EOS is parabolic with respect to isospin asymmetry for a given
interaction in locating the inner edge of neutron star crust.

\begin{acknowledgments}
We would like to thank C. Ducoin, C.M. Ko and S. Kubis for helpful
discussions. This work was supported in part by the National Natural Science
Foundation of China under Grant Nos. 10334020, 10575071, and 10675082, MOE
of China under project NCET-05-0392, Shanghai Rising-Star Program under
Grant No. 06QA14024, the SRF for ROCS, SEM of China, the National Basic
Research Program of China (973 Program) under Contract No. 2007CB815004, the
US National Science Foundation under Grant No. PHY-0652548, PHY-0757839, the
Research Corporation under Award No. 7123 and the Advanced Research Program
of the Texas Coordinating Board of Higher Education Award No.
003565-0004-2007.
\end{acknowledgments}

\end{document}